\def\bold#1{\setbox0=\hbox{$#1$}%
     \kern-.025em\copy0\kern-\wd0
     \kern.05em\copy0\kern-\wd0
     \kern-.025em\raise.0433em\box0 }

\def\slash#1{\setbox0=\hbox{$#1$}#1\hskip-\wd0\dimen0=5pt\advance
       \dimen0 by-\ht0\advance\dimen0 by\dp0\lower0.5\dimen0\hbox
         to\wd0{\hss\sl/\/\hss}}

\documentstyle[12pt]{article}

\newlength{\dinwidth}
\newlength{\dinmargin}
\newcommand{\resection}[1]{\setcounter{equation}{0}\section{#1}}

\def\today{\ifcase\month
\or January\or February\or March\or April\or May
\or June\or July\or August\or September\or October
\or November \or December\fi \space\number\day, \number\year}
\def\draft{\hfill{\it File:\jobname, Draft:\today}}

\begin{document}
%\draft
\vspace*{4cm}
\begin{center}
  \begin{Large}
  \begin{bf}
THE PHYSICS OF THE CHIRAL FERMIONS$^*$\\
  \end{bf}
  \end{Large}
  \vspace{1.5cm}
  \begin{large}
F. Feruglio\\
  \end{large}
Dipartimento di Fisica, Univ.
di Padova\\
I.N.F.N., Sezione di Padova\\
  \vspace{1.5cm}
\end{center}
\begin{quotation}
\begin{center}
  \begin{Large}
  \begin{bf}
  ABSTRACT
  \end{bf}
  \end{Large}
\end{center}
  \vspace{5mm}
\noindent
We review the aspects of chiral gauge theories
related to the violation of the decoupling property.
The case of the top quark is worked out in detail.
The mechanism of anomaly cancellation in the low-energy effective
theory is illustrated in a simple model.
\end{quotation}
\vspace{1.5cm}

\begin{center}
DFPD 94/TH/32\\
May 1994
\end{center}

\vspace{1.0cm}
\begin{center}
\noindent
$^*$ {\it Lectures given at the\\
XVII International School of Theoretical Physics\\
"Standard Model and Beyond '93"\\
Sczcyrk (Poland), September 19-27, 1993.}
\end{center}

%\begin{center}
%DFPD 92/TH/50\\
%September 1992\\
%\vspace{1cm}
%\end{center}
\newpage
\thispagestyle{empty}
\vspace*{7cm}
\begin{quotation}
\tableofcontents
\end{quotation}

\newpage
\setcounter{page}{1}
\def\lq{\left [}
\def\rq{\right ]}
\def\LL{{\cal L}}
\def\DD{{\cal D}}
\def\VV{{\cal V}}
\def\dmu{{\partial_\mu}}
\def\dnu{{\partial_\nu}}
\def\dmua{{\partial^\mu}}
\def\ss{{\left(
         \begin{array}{cc}
         &0\nn\\
         0&\sigma
         \end{array}
         \right)}}
\def\sv{{
         \left(
         \begin{array}{c}
         0\nn\\
         \sigma
         \end{array}
         \right)
         }}
\def\svu{{
         \left(
         \begin{array}{c}
         \sigma\nn\\
         0
         \end{array}
         \right)
         }}
\def\gl{{e^{\dd i\vec\alpha\cdot\frac{\vec\tau}{2}}}}
\def\glm{{e^{\dd -i\vec\alpha\cdot\frac{\vec\tau}{2}}}}
\def\gr{{e^{\dd i\alpha_Y\frac{\tau^3}{2}}}}
\def\grm{{e^{\dd -i\alpha_Y\frac{\tau^3}{2}}}}
\def\da{\frac{\delta\alpha}{\alpha}}
\def\dg{\frac{\delta G_F}{G_F}}
\def\dmz{\frac{\delta M_Z^2}{M_Z^2}}
\def\dmw{\frac{\delta M_W^2}{M_W^2}}

\newcommand{\be}{\begin{equation}}
\newcommand{\ee}{\end{equation}}
\newcommand{\bea}{\begin{eqnarray}}
\newcommand{\eea}{\end{eqnarray}}
\newcommand{\nn}{\nonumber}
\newcommand{\dd}{\displaystyle}

\resection{Introduction}
\indent

These lectures review some low-energy features of gauge theories with
massive chiral fermions.
The standard model (SM), present theory of electroweak interactions,
describes three generations of fermions transforming in chiral
representations of the gauge group $SU(2)_L\otimes U(1)_Y$.
Compared to the electroweak scale defined by the Fermi constant $G_F$,
all fermions are essentially massless, with the exception of the
top quark, whose mass is even larger than the vector boson masses.
This remarkable hierarchy, totally mysterious at the present time,
is accounted for in the theory by a corresponding hierarchy of coupling
constants, which singles out the top Yukawa coupling as
the largest.

Aim of these lectures is to describe the consequences of this
basic fact.
In section 1 we review the decoupling theorem of Appelquist
and Carazzone \cite{AC}. We show how the decoupling property is violated
in the SM with an heavy top quark just because of the assumed relation
between masses and couplings.

Such a violation is not academic. Indeed, as discussed in section 2,
it controls
the pattern of the potentially largest electroweak radiative
corrections and it shows up in the real world with specific
signals. Apart from the existence of lower bounds on
the top mass and the probably imminent
discovery of the top at the Tevatron collider
\footnote{After completion of this review, the CDF collaboration
has announced evidence for signals which, if interpreted as
coming from $t {\bar t}$ production, lead to the estimate:
$m_t=174\pm17~GeV$ \cite{CDF}.}, it is
astonishing how strong and convincing
is the indirect evidence for the top quark
in $B {\bar B}$ oscillations. The often underestimated agreement
of the large set of electroweak precision measurements at LEP and
SLC with the SM expectations is perhaps less striking but it represents
a highly non-trivial fact.

Instead of discussing the full one-loop radiative corrections,
necessary to perform a complete analysis of the LEP/SLC data \cite{fit},
in section 3 we illustrate the pattern of the leading corrections
in the framework of an effective lagrangian. More than a device used
to simplify the discussion of the quantum theory, the effective lagrangian
approach reproduces automatically the infinite set of Ward
identities of $SU(2)_L\otimes U(1)_Y$ \cite{Weinberg},
some of which have revealed so useful
in dealing with leading higher-order computations.

A final lecture is devoted to the mechanism of anomaly cancellation
in the low-energy theory. The gauge invariance of the effective
action, an indispensable requirement, is apparently
broken by the anomalous fermion content of the low-energy spectrum.
This breaking is however repaired by a Wess-Zumino term
whose gauge variation exactly compensates the gauge variation coming from
the classical action.
The independence of the physical amplitudes from the gauge parameter
is thus guaranteed.
\resection{The Decoupling Theorem}

In this section we briefly review the decoupling theorem \cite{AC}.
Consider a field theory with particles of
mass $M$. If the energy at which we perform measurements
is much smaller than $M$, these particles will
affect the predictions of the theory only through their virtual effects.

The decoupling theorem states that, in the limit $M\to \infty$, the
above mentioned effects are unobservable. More precisely, the effects
from heavy particles are either suppressed by inverse powers of  $M$, or
they renormalize parameters of the low-energy theory, that is they
can be absorbed into renormalizations of couplings, masses, wave functions
of the theory obtained by removing the heavy particles.

Examples of theories enjoying the decoupling property are
theories with an exact gauge symmetry, like, for instance,
QED or QCD. The $U(1)$ gauge invariant lagrangian of QED is:
\be
\LL=-\frac{1}{4} F_{\mu\nu} F^{\mu\nu}+
    i {\bar \psi}\gamma^\mu D_\mu \psi
    - M {\bar \psi} \psi~~~,
\ee
where $F_{\mu\nu}=\partial_\mu A_\nu - \partial_\nu A_\mu$ is the
field strength of the photon field $A_\mu$ and the covariant
derivative $D_\mu \psi$ is given by:
\be
D_\mu \psi = (\partial_\mu - i e A_\mu)\psi~~~.
\ee

Suppose that we are interested in the behaviour of the electromagnetic
field $A_\mu$ at energies much smaller than the electron mass $M$.
The first effects potentially affected by $M$ will show up at one-loop order.
%(indeed, due to
%fermion number conservation, one cannot have a tree-level contribution
%with only internal electron lines).
Consider, as the simplest case, the one-loop contribution
$-i\Pi_{\mu\nu}(p)$ to the
photon self-energy. Using dimensional regularization, one obtains:
\be
-i \Pi_{\mu\nu}(p) =
- e^2 \mu^{2\epsilon} \int \frac{d^dk}{(2\pi)^d}
tr(\gamma_\mu\frac{1}{\slash k - \slash p - M}\gamma_\nu
\frac{1}{\slash k - M})~~~.
\ee
where $\epsilon=(4-d)/2$ and $d$ is the space-time dimension.
By introducing the Feynman parametrization, by evaluating
the trace and performing the usual shift in the integration variable,
one has:
\bea
-i \Pi_{\mu\nu}(p) &=&
-  4 e^2 \mu^{2\epsilon} \int \frac{d^dk}{(2\pi)^d}
\int_0^1 dt \frac{1}{(k^2-\Omega)^2}
\Big[\left(\frac{2}{d}-1\right) k^2 g_{\mu\nu} +\nn\\
&+& (p^2 g_{\mu\nu}
-2 p_\mu p_\nu) t(1-t) + M^2 g_{\mu\nu}\Big]~~~,
\eea
where:
\be
\Omega=\Omega(t)=M^2-p^2 t(1-t)
\ee
After the Wick rotation, the integration over the loop variable
gives:
\bea
-i \Pi_{\mu\nu}(p) &=&
-i \frac{8 e^2}{(4\pi)^{d/2}} \mu^{2\epsilon}
(p^2 g_{\mu\nu}- p_\mu p_\nu) \int_0^1 dt~ t(1-t) \Omega^\frac{d-4}{2}
\Gamma\left( \epsilon \right)\nn\\
&=&i \frac{4}{3} \frac{e^2}{(4\pi)^2} (p^2 g_{\mu\nu}- p_\mu p_\nu)
\left[A + 6 \int_0^1 dt~ t(1-t) ln\frac{\Omega(t)}{\mu^2} \right] +...
\eea
where:
\bea
A&=&-\dd\frac{1}{\epsilon}+\gamma_E-ln 4\pi\nn\\
\gamma_E&\simeq&0.577
\eea
and dots in eq. (2.6) stand for terms which vanish in the limit $d \to 4$.
Since we are considering external momenta
much smaller than the electron mass $M$, we can expand the
function $\Omega(t)$ in powers of $p^2/M^2$ and we perform
the (convergent) integration over the Feynman parameter $t$ term by term.
The result is:
\be
-i \Pi_{\mu\nu}(p) =
i \frac{4}{3} \frac{e^2}{(4\pi)^2} (p^2 g_{\mu\nu}- p_\mu p_\nu)
\left[A + ln\frac{M^2}{\mu^2} - \frac{1}{5}\frac{p^2}{M^2} +...\right]
\ee

The previous equation provides a simple example of how the decoupling
property for QED works at one-loop order.
The one-loop self-energy correction (2.8) can be represented by a
set of local terms in an effective low-energy QED lagrangian:
\be
\LL_{eff}=-\frac{1}{4}(1+\delta Z) F_{\mu\nu} F^{\mu\nu}+
c_1 F_{\mu\nu}\Box F^{\mu\nu}+...~~~,
\ee
where dots stand for higher dimensional terms. The coefficients $\delta Z$
and $c_1$ are fixed to reproduce the result given in
eq. (2.8):
\bea
\delta Z &=&
\frac{4}{3} \frac{e^2}{(4\pi)^2} \left[A + ln\frac{M^2}{\mu^2}\right]\nn\\
c_1&=&
\frac{4}{3} \frac{e^2}{(4\pi)^2} \left[- \frac{1}{5 M^2}\right]
\eea

We see that the potentially dangerous
logarithmic dependence on $M$ occurs in the term proportional to
$(p^2 g_{\mu\nu}- p_\mu p_\nu)$ and it is thus absorbed by the wave-
function renormalization of the photon field - $\delta Z$ -
leading to no observable effect.
The next term of the expansion (2.8) cannot be absorbed in a
renormalization of
parameters and is related to an independent operator in the low-energy
effective theory.
However, since the coefficient $c_1$ is inversely proportional to
$M^2$, it vanishes in the limit $M\to \infty$, giving again, in this limit,
no observable effect. This is obviously true also for the remaining terms
in the expansion (2.8).

One can proceed in a completely analogous way with other Green functions,
for instance the four-point photon Green function.
In this way it is easy to check, at one-loop order,
the validity of the decoupling
property for QED or QCD. Appelquist and Carazzone \cite{AC},
extended the proof to all orders in perturbation theory.
\vskip .5truecm
Different from theories possessing an exact gauge symmetry,
theories with spontaneously broken gauge symmetries can be shown
not to necessarily satisfy the decoupling property.
The point is that, whereas in the case of an exact gauge symmetry
mass terms are gauge invariant,
in the spontaneously broken case masses are generated from
interaction terms in the process of symmetry breaking.
The typical mass is of the kind:
\be
M=\lambda <\varphi>
\ee
where $<\varphi>$ is the vacuum expectation value (VEV) of a scalar
field $\varphi$ and $\lambda$ is a dimensionless coupling.
It is clear that, in such a case, the large mass limit can be achieved in
two different ways:

(A) $\lambda$ fixed , $<\varphi>$ large;

(B) $\lambda$ large, $<\varphi>$ fixed.

The first alternative is commonly considered in discussing grand
unified theories (GUTs). In GUTs this choice is suggested by
the physical hierarchy
between the two widely separated VEVs associated to the GUT scale and
to the electroweak one. Other physical situations can however be
described more efficiently by adopting the point of view (B).
Consider for instance the effective lagrangian for low-energy
charged current electroweak processes:
\be
\LL_{CC} = 2\sqrt{2} G_F J^+_\mu {J^-}^\mu~~~,
\ee
where
\be
J^-_\mu=\bar u \gamma_\mu (1-\gamma_5) d +....
\ee
is the total charged current.
In the standard model (SM) of electroweak interactions, the Fermi constant,
$G_F$, is given by:
\bea
\frac{G_F}{\sqrt{2}}&=&\frac{g^2}{8 M_W^2}\nn\\
   &=&\frac{1}{2 v^2}
\eea
where $g$ is the $SU(2)_L$ coupling constant, $v=246~GeV$ is the VEV of
the neutral component of the Higgs doublet. The last expression
of the previous equality represents $G_F$ as a function of $g$ and $v$.
{}From eqs. (2.12) and (2.14),
by considering the large $M_W$ limit according to (A), one obtains
\be
\LL_{CC}\to 0
\ee
in agreement with the decoupling theorem.
However this is not really the case we are interested in. Following (A)
we are considering all SM particles infinitely heavy at the same time.
On the contrary, we have in mind a situation where the external momenta,
of the order of the fermion masses, are much smaller than the $W$
mass:
\be
p^2\simeq m_f^2 \ll M_W^2~~~,
\ee
where $m_f$ is a generic fermion mass.
By denoting with $y_f$ the corresponding Yukawa coupling, the previous
relation implies:
\be
y_f\ll g
\ee
In this case, the large mass limit only reflects the fact that the (light)
fermion Yukawa coupling is much smaller than the $SU(2)_L$ gauge
coupling constant, and therefore it is better represented by the
option (B). As we can see from eq. (2.14), "sending $g$ to infinity"
and keeping $v$ fixed leaves $\LL_{CC}$ invariant and non-vanishing.
The decoupling property is violated.

Notice that, when we speak of large $g$, in the case (B),
we are not saying that $g$ must be much larger than one. Indeed,
the ideal case occurs when $g$, while satisfying the relation
(2.17), still remains smaller than one and the usual
perturbative analysis applies. This is what happens in
the previous example.

A further freedom we have in theories with a spontaneously
broken symmetry is that we can allow a single member in a particle
multiplet to become heavy with respect to the rest of the
spectrum (which is forbidden in the exact case). For instance,
in the SM, we can consider the case of an heavy top quark
whose left-handed component transforms, together with that
of the bottom quark, in an $SU(2)_L$ doublet.

In this case it may seem that the large mass limit is not
compatible with the gauge symmetry one starts with, since one is
removing a member of a representation.
Indeed, to maintain the gauge symmetry in the
light sector, one must embed the light degrees of freedom into
non-linear multiplets, and the symmetry becomes non-linearly
realized \cite{Callan}.
The theory containing the light particles is now non-renormalizable
from the beginning and this, as we shall see in a moment,
can be regarded as a failure of the decoupling property.

To illustrate this point, we consider the Higgs sector of the SM,
described by the $SU(2)_L\otimes U(1)_Y$ invariant lagrangian:
\be
\LL_{H}=
\frac{1}{4}tr(D_\mu H^\dagger D^\mu H)-V(tr(H^\dagger H))
\ee
%\left[\bar q_L H\frac{m_q}{v} q_R + \bar l_L H \frac{m_l}{v} l_R + h.c.\right]
%\eea
The $2$ by $2$ matrix $H$ contains the usual $SU(2)_L$ doublet
of complex scalar fields:
\be
H=
\sqrt{2}\left(\begin{array}{cc}
\varphi^0&-\varphi^+\\
\varphi^-&(\varphi^0)^*
\end{array}\right)
\ee
The covariant derivative $D_\mu H$ is defined by:
\be
D_\mu H=\dmu H - g \hat W_\mu H+g'H\hat B_\mu
\ee
with the $SU(2)_L\otimes U(1)_Y$ gauge fields, $\vec W_\mu$ and $B_\mu$,
embedded in matrices:
\bea
\hat W_\mu&=&\frac{1}{2i}\vec W_\mu\cdot\vec\tau\nn\\
\hat B_\mu&=&\frac{1}{2i} B_\mu\tau^3
\eea
The scalar potential $V$ is given by:
\be
V=\frac{M^2}{8v^2}(\frac{tr(H^\dagger H)}{2}-v^2)^2
\ee
and it depends on $v$, the VEV of $\sqrt{2}\varphi^0$,
and $M$, the mass of the Higgs. By shifting the neutral component according to:
\be
\varphi^0=\frac{h+v+i\chi}{\sqrt{2}}
\ee
one can give the scalar potential the form:
\be
V=\frac{M^2}{8v^2}(h^2+\chi^2+2\varphi^+\varphi^-+2 h v)^2
\ee
%The last terms in eq. (2.18) are Yukawa interactions, to be considered
%later on.

Suppose now that the Higgs is much heavier than the particles
we can excite in a set of physical measurements \cite{Veltman}.
We would like to know if the virtual effects on measurable
quantities due to the heavy Higgs decouple or not, in the sense
specified above. For the moment, we
restrict the analysis to the tree-level approximation.
If we imagine a process with a certain number of
external light particles, it is not so evident, even
in the tree-level approximation, whether the Higgs exchange
will produce negligible effects or not. Indeed, as we can see
from eq. (2.24), there are interaction terms among the Higgs and
the other unphysical scalars which grow as $M^2$, allowing
in principle an overcompensation of the negative powers of $M$
contained in the Higgs propagators.

To fix the ideas, we consider the scattering $\varphi^+\varphi^-\to
\varphi^+\varphi^-$ among unphysical scalars. (As guaranteed by
the so-called equivalence theorem \cite{Cornwall}, this scattering amplitude
is the high-energy approximation to the scattering amplitude
among longitudinally polarized charged vector bosons, and,
to be consistent, we will work in the energy interval
$M_W\ll E\ll M$). At tree level, the amplitude is the sum
of a contact term, plus $s$ and $t$ channel Higgs exchanges. One obtains:
\be
A(\varphi^+\varphi^-\to\varphi^+\varphi^-)=-2i\frac{M^2}{v^2}
-\frac{i M^4}{v^2}\left(\frac{1}{s-M^2}+\frac{1}{t-M^2}\right)
\ee
where we have separately listed the three contributions. While
in the large $M$ limit (at fixed scattering angle)
the leading term, of order $M^2$, cancels, the amplitude, given by:
\be
A(\varphi^+\varphi^-\to\varphi^+\varphi^-)=-i\frac{u}{v^2}+O(1/M^2)~~~,
\ee
is still different from zero
\footnote{Strictly speaking, a failure of the decoupling property
occurs if we are able to observe effects of heavy particles by means
of low-energy tests. This is not transparent from the tree-level
approximation given in eq. (2.26), where the dependence on $M$ vanishes
in the large $M$ limit, while it would be manifest at one-loop level,
when physical observables grow with $ln M$.}.

The general case can be analyzed similarly. We recall that,
summing over all tree-level amplitudes with internal
Higgs lines amounts to:

1. Solve the classical equation of motion for the Higgs field;

2. Substitute back the solution in the original action.

The first step is usually hard to accomplish, because of the non-linearity
of the field equations.
However we are not interested in the full solution of the equations
of motion, but rather in their limit when $M$ is much larger
than the energy and/or other mass parameters.
To this end we parametrize the scalar multiplet $H$ as follows:
\be
H=\sigma U=\sigma e^{\dd i\frac{\vec\xi\cdot\vec\tau}{v}}
\ee
where the fields $\vec\xi=(\xi^1,\xi^2,\xi^3)$ describe the would-be Goldstone
bosons.
The lagrangian for the scalar sector reads:
\bea
\LL&=&\frac{1}{2}\dmu\sigma\dmua\sigma-
           \frac{M^2}{8 v^2}(\sigma^2-v^2)^2+\nn\\
           &+&\frac{\sigma^2}{4}tr(D_\mu U^\dagger D^\mu U)
%-\nn\\
%           &-&\left[\left(\frac{\sigma}{v}\right)\bar q_L U m_q q_R+
%             \left(\frac{\sigma}{v}\right)\bar l_L U m_l l_R + h.c.\right]
\eea
The Higgs degree of freedom is now described by the field $\sigma$.
In the large $M$ limit, the solution of the equation of motion
for $\sigma$ is simply:
\be
\sigma=v
\ee
since for large $M$ the action is dominated by the scalar potential.
Plugging back this solution in the lagrangian of eq. (2.28), one finds
\cite{Appel}:
\be
\LL(M=\infty)=\frac{v^2}{4}tr(D_\mu U^\dagger D^\mu U)
%           &-&\left[\bar q_L U m_q q_R+
%             \bar l_L U m_l l_R + h.c.\right]
\ee
This is the lagrangian for a (gauged) non-linear $\sigma$-model \cite{Gursey}.
It contains an infinite set of operators depending on the would-be Goldstone
fields $\vec \xi$, which makes it a non-renormalizable theory.
No field redefinition can turn it into a renormalizable lagrangian.
It represents the sum of all one-particle irreducible tree diagrams
with infinitely heavy Higgs internal lines. It is easy to check
that the amplitude for $\xi^+\xi^-\to\xi^+\xi^-$ derived from it
coincides with that evaluated before in eq. (2.26).

The Higgs particle was originally a member of an $SU(2)_L$
doublet (see eq. (2.19)). To separate it from the rest of the
doublet consistently with the gauge invariance,
we have performed the field transformation in eq. (2.27).
This leads to a low-energy theory for the would-be
Goldstone modes $\vec\xi$ with the electroweak symmetry
realized non-linearly \cite{Appel,Longhi1,Longhi2,Her}.
The occurrence of infinitely many
higher-dimensional operators with coefficients not suppressed
by inverse power of $M$ signals the failure of the decoupling property.

This failure persists at the quantum level. However, owing to a remarkable
property of the SM, at one-loop order the dependence of the generic
physical observable upon the Higgs mass is only logarithmic
(screening theorem) \cite{Veltman}.
Power-like effects are possible, but only at
higher orders. In physical terms this means that detection of the
Higgs through its virtual effects will not be easy. This screening
effect is strictly related to the minimal structure of the SM
and power-like dependencies can be generated in modest extensions
of the SM as, for instance, in models containing two scalar doublets
\cite{power}.
\vskip .5truecm

Chiral fermions,
that is fermions whose left and right-handed components
transform according to inequivalent representations of the gauge group
(contrary to vector-like fermions),
provide other examples
of violation of the decoupling property \cite{D'Hoker}. Chiral fermions
do not admit gauge invariant
mass terms and their masses are generated via the spontaneous breaking
of the gauge symmetry, from Yukawa interactions.
When a chiral fermion is made heavier than the other matter fields
by a relatively large Yukawa coupling, its effects at low energies
do not decouple.
The mass suppression associated
to the propagator can be compensated by the mass enhancement
provided by vertices with an overall non-vanishing effect.
This mechanism is well exemplified in the SM by the top quark.
The top is by far the heaviest of
the known fermions. The top Yukawa coupling - $y_t$ - is
of order one (about .7 for $m_t=174~GeV$),
much larger than the other Yukawa couplings and comparable
with the $SU(2)_L$ gauge coupling $g$ ($g\simeq .65$).
In the ideal case where we could neglect $g$ and $g'$, the top quark
would provide extremely clean signals of breakdown of the decoupling
property, ordered only by powers of $y_t$ and, as we shall see,
easy to compute. In the real world $y_t$ and the
gauge couplings $g$ and $g'$ are of the same order and, depending on the
physical observable considered, we expect significant corrections
to the previous, ideal case.

Consider first
the low-energy effective action for an
heavy top quark, in the tree-level approximation.
As we have seen previously, we have to solve the classical
equations of motion for the top quark, in the limit $E\ll m_t$.
The lagrangian for the quarks reads:
\bea
\LL_q&=&i\bar q_L\gamma^\mu D_\mu q_L +
         i\bar q_R\gamma^\mu D_\mu q_R -\nn\\
&-&
\left[\bar q_L H\frac{m_q}{v} q_R + h.c.\right]
\eea
We have put both the up and the down type quarks in a single multiplet $q$
and indices in the generation space are understood.
The covariant derivatives acting on the left and right-handed quarks
are defined below:
\bea
D_\mu q_L&=&(\dmu- g\hat W_\mu - g'\hat B^{(L)}_\mu) q_L\\
D_\mu q_R&=&(\dmu- g'\hat B^{(R)}_\mu) q_R
\eea
The combinations $\hat B^{(L,R)}_\mu$ are given by:
\bea
\hat B^{(L)}_\mu&=&\frac{1}{6i} B_\mu\\
\hat B^{(R)}_\mu&=&\frac{1}{2i}(\tau^3+\frac{1}{3}) B_\mu
\eea
In the large $m_t$ limit,
the only relevant term of the action is the Yukawa coupling
of eq. (2.31), which, more explicitly, reads:
\be
\LL_Y=-\frac{\sigma}{v}
\left(\bar u^0_L \bar d^0_L\right) U
\left(\begin{array}{cc}m_u&0\\ 0&m_d\end{array}\right)
\left(\begin{array}{c}u^0_R\\d^0_R\end{array}\right) + ~h.c.
\ee
where $m_u$ and $m_d$ are the 3 by 3 quark mass matrices in the up and
down sectors, respectively. Diagonal mass matrices $m^D_u$, $m^D_d$
and mass eigenstates $u_{L,R}$, $d_{L,R}$ are introduced
via a bi-unitary transformation:
\bea
m^D_u&=&\VV^u_L m_u {\VV^u_R}^\dagger\nn\\
m^D_d&=&\VV^d_L m_d {\VV^d_R}^\dagger
\eea
\bea
u_{L,R}=\VV^u_{L,R} u^0_{L,R}\nn\\
d_{L,R}=\VV^d_{L,R} d^0_{L,R}
\eea
The Yukawa lagrangian, in terms of mass eigenvalues and eigenstates,
reads:
\be
\LL_Y=-\frac{\sigma}{v}
\left(\bar u_L \bar d^W_L\right) U
\left(\begin{array}{cc}m^D_u&0\\ 0&V_{CKM} m^D_d {V_{CKM}}\dagger
\end{array}\right)
\left(\begin{array}{c}u_R\\d^W_R\end{array}\right) + ~h.c.
\ee
where $V_{CKM}=\VV^d_L {\VV^u_L}^\dagger$ is the Cabibbo-Kobayashi-
Maskawa (CKM) mixing matrix and $d^W_{L,R}=V_{CKM} d_{L,R}$.
If we assume that all fermion masses but the top one are
zero, we can isolate from the previous equation the top sector:
\be
\LL_Y=-\frac{\sigma}{v}
\left(\bar t_L \bar b^W_L\right) U
\left(\begin{array}{cc}m_t&0\\ 0&0
\end{array}\right)
\left(\begin{array}{c}t_R\\b^W_R\end{array}\right) + ~h.c.
\ee
Notice that the top couples to the "weak" bottom combination:
\be
b^W=V_{td} d+V_{ts} s + V_{tb} b~~~,
\ee
mainly made of the physical bottom quark. From the unitarity
of $V_{CKM}$ and the measured entries of the $V_{CKM}$ matrix,
one has the following 90\% CL bounds \cite{unit}:
\bea
0.003 &\le V_{td} &\le 0.018\nn\\
0.030 &\le V_{ts} &\le 0.054\nn\\
0.9985 &\le V_{tb} &\le 0.9995
\eea

The equations of motion for the top quark, in the large $m_t$ limit,
decouple into two separate equations for the left and for the
right-handed components:
\be
U
\left(\begin{array}{cc}1&0\\ 0&0
\end{array}\right)
\left(\begin{array}{c}t_R\\b^W_R\end{array}\right) = 0
\ee
\be
\left(\begin{array}{cc}1&0\\ 0&0
\end{array}\right) U^\dagger
\left(\begin{array}{c}t_L\\b^W_L\end{array}\right) = 0
\ee
whose solutions are:
\be
t_R=0
\ee
\be
\left(\begin{array}{c}t_L\\b^W_L\end{array}\right) =
U \left(\begin{array}{c} 0 \\{b^W_L}'\end{array}\right)
\ee
Inserting back these solutions in the lagrangian $\LL_q$ of eq. (2.31),
one obtains \cite{FMM}:
\bea
\LL_q &=& i {\bar b}\gamma^\mu \partial_\mu b + \frac{g'}{3} B_\mu
          {\bar b}\gamma^\mu b\nn\\
&-&\frac{i}{2}{\bar b}^W_L \gamma^\mu b^W_L tr(T V_\mu) +...
\eea
where:
\be
T=U \tau^3 U^\dagger~~~,
\ee
\be
V_\mu=D_\mu U \cdot U^\dagger~~~,
\ee
and dots stand for the remaining light quarks.
The gauge invariant combination:
\be
-\frac{i}{2}{\bar b}^W_L \gamma^\mu b^W_L tr(T V_\mu)=
{\bar b}^W_L \gamma^\mu b^W_L\left[\frac{g}{2\cos\theta} Z_\mu
+ \frac{\partial_\mu \xi^3}{v}+...\right]
\ee
contains infinitely many terms representing interactions
of the $V-A$ bottom current with gauge and would-be Goldstone bosons.
This is precisely the term which represents the non-vanishing
sum of all tree diagrams with internal, heavy top lines.
(The apparent flavour changing neutral current contained in the first
term of the expansion - $Z_\mu {\bar b}^W_L \gamma^\mu b^W_L$ -
is cancelled by the contributions from the light fermions.)

As in the case of the Higgs boson previously discussed, we have been
able to remove the heavy top quark from the low-energy theory
at the expense of introducing an infinite set of higher dimensional
terms, not suppressed by inverse powers of $m_t$. On the contrary,
as we shall see in the next sections, at one-loop
measurable quantities start depending upon the square of
the top mass, leading to new physical effects and
making non-trivial the agreement between the
SM predictions and the available data from precision experiments.

\resection{Non-decoupling in the Real World}

In this section we summarize the phenomenological relevance of
the one-loop electroweak corrections due to the top quark.

The important role played by the top quark in the analysis of
neutral current data and the possibility of deriving a bound
on $m_t$ from low-energy measurements was recognized quite soon
\cite{AmCo}.
\vskip .5truecm

An important development took place in 1987, with the discovery
of $B^0-{\bar B^0}$ oscillations. It became then clear that
the top quark was much heavier than previously expected.

The first evidence of $B^0-{\bar B^0}$ oscillations was
found by the UA1 collaboration \cite{UA1}. The theoretical interpretation
of their data was however difficult, since $B^0_d$ and $B^0_s$
were produced in an essentially unknown mixture.
Later on the signal was confirmed by the ARGUS \cite{ARGUS} and CLEO
\cite{CLEO} collaborations, who
found evidence for equal sign dileptons in the decay of $\Upsilon(4s)$.
The $\Upsilon(4s)$ resonance, through its decay into a $B^0_d-{\bar B}^0_d$
pair, gives rise to a final state containing two charged leptons:
$l^+$, coming from $B^0_d$ and $l^-$, coming from ${\bar B}^0_d$.
If a $B^0_d-{\bar B^0}_d$ mixing is allowed, then, sometimes
the $B^0_d$ decay produces $l^-$,
the ${\bar B^0}_d$ gives rise to $l^+$ and, in a fraction of the events,
one will find equal sign dileptons.
The relevant parameter is the ratio of equal sign to opposite sign leptons:
\be
r_d=\frac{N(l^+l^+) + N(l^-l^-)}{N(l^+l^-)}\big|_{\Upsilon(4s)}
\ee
which experimentally is given by:
\be
r_d=0.17\pm0.10
\ee
More recently also the LEP collaborations have found an excess of
like sign lepton pairs in $e^+e^-\to l^\pm l^\pm+X$, coming
both from $B^0_d$ and $B^0_s$. Moreover they were able to detect
the predicted time dependence of the $B^0_d$ oscillations \cite{LEPbb}.

The $B^0_d - {\bar B^0}_d$ system is described by the two-
dimensional effective hamiltonian:
\be
H=
\left(
\begin{array}{cc}
M&M_{12}\nn\\
M^*_{12}&M
\end{array}
\right)-
\frac{i}{2}
\left(
\begin{array}{cc}
\Gamma&\Gamma_{12}\nn\\
\Gamma^*_{12}&\Gamma
\end{array}
\right)
\ee
written in the base $(B^0_d {\bar B^0}_d)$.
In the $B$ system $\Gamma_{12}$ is approximately zero and the
two eigenstates of the hamiltonian have essentially the same
width $\Gamma$, which can be extracted directly from the measured
lifetime $\tau_B$. On the other hand, the mass difference $\Delta M$
between the two eigenvalues is given by:
\be
\Delta M=2\vert M_{12}\vert
\ee
Introducing the two parameters:
\bea
x&=&\frac{\Delta M}{\Gamma}\nn\\
y&=&\frac{\Delta\Gamma}{2\Gamma}
\eea
one obtains:
\be
r_d=\frac{x^2+y^2}{2+x^2-y^2}\simeq \frac{x^2}{2+x^2}
\ee
The $x$ parameter can be estimated from
the $\Delta B=2$ non-leptonic effective hamiltonian,
which in the SM arises as a result of a second order weak interaction:
\bea
H(\Delta B=2)&=&\frac{G_F}{16\pi^2}(V^*_{tb} V_{td})^2 m_t^2 f(\frac{m_t^2}
{M_W^2})\eta\cdot\nn\\
&&\cdot{\bar b}\gamma^\mu (1-\gamma_5) d {\bar b}\gamma_\mu (1-\gamma_5) d
\eea
where $\eta$ is a factor of order one which accounts for the QCD
corrections and $f(z)$ is a slowly varying function of $z$:
\be
f(z)=\frac{1}{4} +\frac{9}{4(1-z)}-
\frac{3}{2(1-z)^2}+\frac{3}{2}\frac{z^2 ln z}{(z-1)^3}
\ee
with $f(1)=3/4$ and $f(\infty)=1/4$.
{}From $H(\Delta B=2)$ and eq. (3.4) one can derive $\Delta M$:
\be
\Delta M=2\vert<{\bar B^0}_d\vert H(\Delta B=2) \vert B^0_d>\vert
\ee
This requires the computation of the hadronic matrix element:
\be
<{\bar B^0}_d\vert
{\bar b}\gamma^\mu (1-\gamma_5) d {\bar b}\gamma_\mu (1-\gamma_5) d
\vert B^0_d>=\frac{4}{3} B_B f_B^2 m_B
\ee
where $m_B$ is the $B$ meson mass, $f_B$ its decay constant
and $B_B$ parametrizes a possible departure from
the so-called vacuum saturation approximation in which $B_B=1$.
{}From equations (3.5), (3.7), (3.9) and (3.10)
one obtains:
\be
x=\frac{G_F^2}{6\pi^2} m_t^2 f(\frac{m_t^2}{M_W^2}) B_B f_B^2 m_B \tau_B
\vert V^*_{tb} V_{td}\vert ^2 \eta
\ee
Notice the leading quadratic dependence of $x$ on the top mass,
coming from the box diagram which, in the SM, gives rise to $H(\Delta B=2)$.
We shall come back to this point in section 4.
Apart from the top mass, $x$ also depends on the hadronic
parameter $\sqrt{B_B} f_B$ and the CKM combination
$\vert V^*_{tb} V_{td}\vert$.
On the former quantity, we have estimates from the QCD sum rules approach
and from the lattice, from which we expect $\sqrt{B_B} f_B$ in the range
$100-300~MeV$ \cite{bfact}. The CKM angles involving the top quark are
presently
unknown, but restrictions on them can be derived from the unitarity
of the CKM matrix $V$ assuming that only three generations
are present. One finds \cite{unit}:
\be
0.003\le \vert V^*_{tb} V_{td}\vert \le 0.018
\ee
Finally $\eta=0.78-0.85$ \cite{eta}. Putting everything together \cite{Lusi},
one realizes that, even pushing all the unknown quantities to the
extreme upper limit compatible with the present bounds or our
theoretical understanding,
a large value for $m_t$ ($m_t \ge 50~GeV$) is required to have consistence with
the experimental value of $r_d$.
\vskip .5truecm

Other observables affected by potentially large $m_t$
corrections are those related to the electroweak precision measurements
done at LEP/SLC. With the exception of the partial width of the Z into
$\bar b b$, which will be discussed at the end of this section,
the leading top quark effects, at one-loop level, are
dominated by the gauge bosons self-energy corrections.
To count the number of independent parameters occurring in this sector
\cite{AB},
we start from the usual definition:
\be
-i\Pi_{ij}^{\mu\nu}(p)=-i\left[\Pi_{ij}(p^2)g^{\mu\nu}+
\left(p^\mu p^\nu~~ terms\right)\right]
\ee
where $-i\Pi_{ij}^{\mu\nu}(p)$ denote the set of self-energies
for the gauge boson fields.
The indices $i,j$ can take the values $0$ (for the field $B$) and
$1,2,3$ (for the fields $W^i$),
or, alternatively, the values $\gamma,~Z,~W$.
{}From now on we will
discard the irrelevant terms proportional to $p^\mu p^\nu$.
Furthermore, we make a Taylor expansion of the top contribution
to the scalar function $\Pi_{ij}(p^2)$, around the point $p^2=0$:
\be
\Pi_{ij}(p^2)=A_{ij}+p^2F_{ij}+...
\ee

This expansion, meaningful for
$p^2\ll m_t^2$ contains real coefficients $A_{ij}$, $F_{ij}$, etc.
Moreover, since $\Pi_{ij}(p^2)$ has dimension two in units of mass,
it is reasonable to neglect the dots in eq. (3.14), representing
terms suppressed by positive powers of $(p^2/m_t^2)$.

As a consequence of the exact electromagnetic gauge invariance,
we have $A_{\gamma\gamma}=A_{\gamma Z}=0$. (More precisely,
the fermionic contribution to $A_{\gamma Z}$
vanishes, and the bosonic one is zero in the unitary gauge.)
Then
we are left with the six independent coefficients $A_{ZZ}$, $A_{WW}$,
$F_{\gamma\gamma}$, $F_{\gamma Z}$, $F_{ZZ}$, $F_{WW}$, carrying the
main dependence on $m_t$.
Three combinations of them are however unobservable, being
related to the fundamental
constants of the electroweak theory: the electromagnetic fine structure
constant $\alpha$, the Fermi constant $G_F$ and the mass of the
$Z$ gauge vector boson $M_Z$.
Indeed, the quantum corrections induced by the gauge
vector boson self-energies provide the following
shifts in the fundamental constants
\footnote{In a general analysis of the one-loop corrections
one should also include
in $\delta G_F$ contributions coming from boxes, vertices and
fermion self-energies. Similarly, the right-hand side of
eq. (3.17) would read $-\Pi_{ZZ}(M_Z^2)/M_Z^2$ \cite{Barb,Yellow}. However,
since here we are only interested
in the dependence upon the top quark mass, the additional contributions
can be neglected.}:
\bea
\frac{\delta\alpha}{\alpha}&=&-F_{\gamma\gamma}\\
\frac{\delta G_F}{G_F}&=&\frac{A_{WW}}{M_W^2}\\
\frac{\delta M_Z^2}{M_Z^2}&=&-\left(\frac{A_{ZZ}}{M_Z^2}+F_{ZZ}\right)
\eea

We conclude that, in our approximation,
the parameters carrying in the SM the leading top quark dependence
are three combinations among the six coefficients $A_{ZZ},~A_{WW},
{}~F_{\gamma\gamma},~F_{\gamma Z},~F_{ZZ},~F_{WW}$.
These combinations can be identified by looking at the radiative corrections
for three independent physical observables, which we choose as
the ratio of the
gauge boson masses $M_W/M_Z$, the forward-backward
asymmetry $A_{FB}^\mu$ in $e^+e^-\to
\mu^+\mu^-$ at the $Z$ peak and the partial width of the $Z$ into charged
leptons, $\Gamma_l$.
\vskip .5truecm
\newpage

\noindent
$\bullet$
$\dd\frac{M_W}{M_Z}$

We trade ${M_W}/{M_Z}$ for the observable $\Delta r_W$ defined as
follows:
\be
\left(\frac{M_W}{M_Z}\right)^2=\frac{1}{2}+
\sqrt{\frac{1}{4}-\frac{\mu^2}{M_Z^2 (1-\Delta r_W)}}
\ee
where:
\be
\mu^2=\frac{\pi\alpha(M_Z^2)}{\sqrt{2} G_F}=(38.454~GeV)^2
\ee
One finds:
\bea
\Delta r_W&=&-\frac{\cos^2\theta}{\sin^2\theta}\left(\frac{A_{ZZ}}{M_Z^2}-
\frac{A_{WW}}{M_W^2}\right)+\nn\\
&+&\frac{\cos 2\theta}{\sin^2\theta}\left(F_{WW}-F_{33}\right)+\nn\\
&+&2\frac{\cos\theta}{\sin\theta} F_{30}
\eea
\vskip .5truecm

\noindent
$\bullet$
${\bf A_{FB}^\mu},~~~ {\bf \Gamma_l}$

By forward-backward asymmetry at the peak we mean
the quantity quoted by the LEP experiments, which is corrected
for all QED effects, including initial and final state radiation and also
for the effect of the imaginary part of the photon vacuum polarization diagram.
The partial width of $Z$ into charged leptons is inclusive of
photon emissions: $\Gamma_l=\Gamma(Z->l {\bar l} + photons)$.

Also in this case we proceed through a series of definitions
inspired to the lowest order relations:
\be
A_{FB}^\mu (p^2=M_Z^2)=3\left(\frac{g_V g_A}{g_V^2+g_A^2}\right)^2
\ee
\be
\Gamma_l=\dd\frac{G_F M_Z^3}{6\pi\sqrt{2}}
\left(g_V^2+g_A^2\right)(1+\frac{3 \alpha}{4\pi})
\ee
\bea
g_A&=&\frac{1}{2} (1 + \frac{\Delta\rho}{2})\\
g_V/g_A&=&-1+4\sin^2\hat\theta
\eea
\be
\sin^2\hat\theta=(1+\Delta k)\sin^2\tilde\theta
\ee
\bea
\sin^2\tilde\theta&=&\frac{1}{2}-
\sqrt{\frac{1}{4}-\frac{\mu^2}{M_Z^2}}\nn\\
&=&0.23118~~~~({\rm for}~~M_Z=91.187~GeV)
\eea

With these definitions, the knowledge of $A_{FB}^\mu$, which depends
only on the ratio $g_V/g_A$ is
equivalent to that of the parameter $\Delta k$, given in eq. (3.25).
On the other hand the parameter $\Delta\rho$, entering the definition of the
$Z$ coupling to charged leptons as an overall factor, is fixed by
$\Gamma_l$. One finds:
\bea
\Delta k=&-&\frac{\cos^2\theta}{\cos 2\theta}
\left(\dd\frac{A_{ZZ}}{M_Z^2}-\dd\frac{A_{WW}}{M_W^2}\right)\nn\\
&+&\dd\frac{1}{\cos 2 \theta}\dd\frac{\cos\theta}{\sin\theta} F_{30}
\eea
\be
\Delta\rho=\dd\frac{A_{ZZ}}{M_Z^2}-\dd\frac{A_{WW}}{M_W^2}
\ee

Indeed the whole set of self-energy corrections can be accounted for
by an effective neutral current hamiltonian given by:
\be
H_{NC}=\left(4 \sqrt{2} G_F M_Z^2\right)^\frac{1}{2} (1+\frac{\Delta\rho}{2})
\left[J^\mu_{3L}-(1+\Delta k) \sin^2\tilde\theta J^\mu_{em}\right]Z_\mu
\ee

So far we have selected three physical quantities
in order to isolate their leading dependence upon the top quark mass.
By looking at the expressions obtained for the
quantities $\Delta r_W$, $\Delta k$ and $\Delta\rho$,
we recognize that they are functions of the following three
combinations of self-energy corrections
\cite{Peskin,AB,ABJ}:
\bea
\epsilon_1&=&\dd\frac{A_{ZZ}}{M_Z^2}-\dd\frac{A_{WW}}{M_W^2}\nn\\
\epsilon_2&=&F_{WW}-F_{33}\nn\\
\epsilon_3&=&\dd\frac{\cos\theta}{\sin\theta} F_{30}
\eea
We can summarize the results as follows:
\bea
\Delta r_W&=&-\dd\frac{\cos^2\theta}{\sin^2\theta}\epsilon_1
+\dd\frac{\cos 2\theta}{\sin^2\theta}\epsilon_2
+2\epsilon_3\nn\\
\Delta k&=&-\dd\frac{\cos^2\theta}{\cos 2 \theta}\epsilon_1+
\dd\frac{1}{\cos 2\theta}\epsilon_3\nn\\
\Delta\rho&=&\epsilon_1
\eea
We remind that the relationship exhibited by eqs. (3.30-31)
reflects the fact that in the SM most of the top quark
contribution to the considered observables is contained
in the vacuum polarization functions of the vector gauge bosons,
suitably expanded as in eq. (3.14).
The more general dependence of $\Delta r_W$, $\Delta k$ and $\Delta\rho$
on the SM radiatiative corrections can be easily derived
along lines similar to those followed here, and it would include
vertex, box and fermion self-energy corrections as well
\cite{Barb}. The latter
do not contain any further significant dependence on $m_t$.

Within the SM, the combinations in eq. (3.30) have the following
asymptotic dependence on $m_t$:
\bea
\epsilon_1&=&\frac{3 G_F m_t^2}{8 \pi^2 \sqrt{2}}+...\\
\epsilon_2&=&-\frac{G_F M_W^2}{2 \pi^2 \sqrt{2}}ln(\frac{m_t}{M_Z})+...\\
\epsilon_3&=&-\frac{G_F M_W^2}{6 \pi^2 \sqrt{2}}ln(\frac{m_t}{M_Z})+...
\eea
Notice that the potentially largest top quark correction, namely the one
quadratic in $m_t$, appears only in $\epsilon_1$, while in $\epsilon_2$ and
$\epsilon_3$ the dependence on $m_t$ is only logarithmic.

Eq. (3.31) is the starting point
of the so-called non-standard analysis of the electroweak data
\cite{ABJ,ABC}.
Indeed, forgetting about the way eq. (3.31) was derived,
one can take it as the {\it definition} of the $\epsilon$
parameters, which become true physical observables,
with the advantage that the strongest dependence on $m_t$ has been
confined in $\epsilon_1$. The inclusion of a larger set of experimental
data, to provide further information on the $\epsilon$ parameters,
demands some further assumptions, which can be ordered according to
an increasing
amount of model dependence. This offers a common ground
to compare various theoretical
frameworks (SM \cite{ABJ,ABC},
minimal supersymmetric standard model \cite{ABC2},
extended gauge models \cite{Zprime}, ...).
{}From the experimental values $M_W/M_Z=0.8798\pm0.0028$,
$A_{FB}^l=0.0170\pm0.0016$ and $\Gamma_l=83.975\pm0.20~MeV$,
one finds \cite{Altarelli}:
\bea
\epsilon_1&=&(0.42\pm0.24)\cdot10^{-2}\nn\\
\epsilon_2&=&(-0.25\pm0.56)\cdot10^{-2}\nn\\
\epsilon_3&=&(0.35\pm0.31)\cdot10^{-2}
\eea
\vskip .5truecm

If vacuum polarization corrections were always dominating, at least for
the part concerning the top dependence, then, from the effective
hamiltonian $H_{NC}$ in eq. (3.29), one would conclude that,
for all flavours $f$,
the partial width $\Gamma_f$ of the $Z$ boson into $f {\bar f}$
is given by
\footnote{apart from QED and QCD corrections}:
\be
\Gamma_f=N_C^f \frac{G_F M_Z^3}{6 \pi \sqrt{2}}
         \left[(g_V^f)^2+(g_A^f)^2\right]
\ee
with $g_V^f$ and $g_A^f$ given by:
\bea
g_V^f&=&(1+\frac{\Delta\rho}{2})(T_{3L}^f-2 Q_{em}^f \sin^2\hat\theta)\\
g_A^f&=&-(1+\frac{\Delta\rho}{2})T_{3L}^f
\eea
Then the effective fermionic couplings of the $Z$ boson would
be characterized by
universal, flavour-independent, corrections: $\Delta\rho$
and $\Delta k$.
However, because of the occurrence of vertex corrections,
this conclusion is not true,
not even for the top contribution, and the largest
violation of eq. (3.37-38) takes place for $f=b$.
For $Z$ decaying into $b {\bar b}$, besides the vacuum polarization effects
one should also take into account the vertex corrections and the fermion
self-energy corrections, where the exchange of charged unphysical
scalars gives rise to additional terms quadratic in $m_t$ \cite{Zbb}.
In this case one has still the expression of eq. (3.36) for the partial
width $\Gamma_b$, but $g_A^b$ and $g_V^b$ are replaced by:
\bea
g_A^b&=&\frac{1}{2}(1+\frac{\Delta\rho}{2})(1+\epsilon_b)\\
\frac{g_V^b}{g_A^b}&=&
\dd\frac
{-1+\frac{4}{3}\sin^2\hat\theta-\epsilon_b}
{1+\epsilon_b}
\eea
The new corrections are isolated in the parameter $\epsilon_b$,
whose asymptotic dependence on the top quark mass reads:
\be
\epsilon_b=-\frac{G_F m_t^2}{4 \pi^2 \sqrt{2}}+...\\
\ee
The additional corrections
can be derived from an effective hamiltonian of the form:
\be
H_b=-\frac{1}{2}\epsilon_b
\left(4 \sqrt{2} G_F M_Z^2\right)^\frac{1}{2}
Z^\mu {\bar b}_L \gamma_\mu b_L
\ee
The presence of the $\epsilon_b$ term in $\Gamma_b$, quadratic in $m_t$,
singles out this partial width as a particularly interesting quantity,
whose peculiar dependence on $m_t$ is potentially able to provide
additional and independent information on the top quark.

{}From the present value $\Gamma_b=385.3\pm3.9~MeV$ and by removing from
$\Gamma_b$ the QCD correction, one obtains
\cite{Altarelli}:
\be
\epsilon_b=(0.46\pm0.45)\cdot10^{-2}
\ee

\resection{An Effective Lagrangian for the Heavy Top Quark}

As examples of
violation of the decoupling property, we have seen that the heaviness
of the top quark shows up in three independent effects:
the $B^0-{\bar{B^0}}$ oscillations; the vacuum polarization
corrections (in particular the $\Delta\rho$ parameter)
affecting all LEP/SLC observables and the $M_W/M_Z$ mass ratio;
the non-universal correction
of the $Zb{\bar b}$ vertex detectable through the measure
of the $\Gamma_b$ partial width.
The leading part of these effects can be described by
the effective hamiltonians
given in eqs. (3.7), (3.29) and (3.42). In this section we will discuss
the general structure of the effective lagrangian which reproduces,
at one-loop order, the above mentioned effects \cite{FMM,Steger}.

To start with we observe that one-loop results are
not correctly reproduced by the effective lagrangian
we have derived in the large $m_t$ limit in section 2,
namely ${\cal L}_{cl}$ given by the sum of $ {\cal L}_q$ of eq. (2.47),
${\cal L}_H$ of eq. (2.18) and the terms for the
other light fermions and the
gauge vector bosons.
This has to do with the fact that ${\cal L}_{cl}$ was obtained
via a classical limit, corresponding to the sum of all
tree-level diagrams containing heavy top quark lines.
To deal correctly with the one-loop computation, we must
first perform the (regularized) loop integration and
subsequently take the large $m_t$ limit. In general this leads
to a result which is a divergent function of the ultraviolet
cut-off.
For this reason the opposite way,
namely first taking the large $m_t$ limit - which amounts to use
${\cal L}_{cl}$ - and then performing the loop integration,
generally leads to a different result.
In formulae:
\be
\lim_{m_t\to\infty} \int dk~ F_\Lambda (k,p) = \int dk~\lim_{m_t\to\infty}
F_\Lambda (k,p) + \Delta (p)
\ee
where $p$ stands for a collection of external momenta, $k$ is the
loop variable, $\Lambda$ an ultraviolet cut-off. The function
$\Delta (p)$ represents the $O(\slash h)$ correction
which we should add to the result obtained working with ${\cal L}_{cl}$,
to correctly reproduce the one-loop result in the large $m_t$ limit.

The interesting fact is that $\Delta (p)$ can be represented as
an effective lagrangian.
For external momenta lighter than $m_t$ the integrals on both sides
of the eq. (4.1) have equal imaginary parts in all possible channels,
and therefore the function $\Delta (p)$ is an analytic
function of the variables $p$. If we consider its expansion in
$p^2$, for dimensional reasons, there will be only a finite number of terms
not vanishing in the large $m_t$ limit. Moreover, for amplitudes with
a sufficiently large number of external legs the loop integral
is convergent, the $m_t \to \infty$ limit and the loop integral commute
and $\Delta (p)=0$.
We conclude that at one-loop order the correct results of the
large $m_t$ limit are reproduced
by adding to ${\cal L}_{cl}$ a finite number of local terms,
which we collectively denote by $\Delta {\cal L}$.
The low-energy theory, in the $m_t\to \infty$ limit and to one-loop
accuracy, is described by the effective lagrangian:
\be
{\cal L}_{eff}={\cal L}_{cl} + \Delta {\cal L}
\ee
The term $\Delta {\cal L}$ is further restricted by the symmetry
of the low-energy theory. For the moment we require $\Delta{\cal L}$
to be $SU(2)_L\otimes U(1)_Y$ invariant (see however the next section).
In section 2 we have already made use of non-linear
realizations of the $SU(2)_L\otimes U(1)_Y$ symmetry to describe
${\cal L}_{cl}$ \cite{Appel,Longhi1,Longhi2,Her}.
The non-linear realization naturally provides a low-energy
expansion, ordered by the number of derivatives acting on the
light fields. In our case such an expansion should contain
terms of order $p^4$. Indeed we must take into
account at least the terms $F_{ij}$ introduced in the analysis
of vacuum polarization (see eq. (3.14)). These are terms of order
$p^2$ containing two gauge vector bosons, which gauge invariance relates
to terms of order $p^4$ \cite{DeRujula}.
In addition to $SU(2)_L\otimes U(1)_Y$ gauge invariance we
will also ask for $CP$ invariance.
We list below the invariant operators which are relevant to our discussion,
containing up to four derivatives and built out the
gauge vector bosons $W$, $Z$, $A$ and the would be Goldstone bosons
$\vec\xi~~$ \footnote{For a complete list, see, for instance, ref.
\cite{Appel2}.}:
\bea
\LL_0&=&\dd\frac{v^2}{4}[tr(TV_\mu)]^2\nn\\
\LL_1&=&i\dd\frac{gg'}{2}B_{\mu\nu}tr(T\hat W^{\mu\nu})\nn\\
\LL_8&=&\dd\frac{g^2}{4}[tr(T\hat W_{\mu\nu})]^2
%\nn\\
%\LL_{13}&=&\dd\frac{1}{2}[tr(T \DD_\mu V_\nu)]^2\nn\\
\eea
These operators contribute to $\Delta {\cal L}$ in eq. (4.2)
through the term:
\be
(\Delta{\cal L})_B= a_0 \LL_0 + a_1 \LL_1 + a_8 \LL_8
%+ a_{13} \LL_{13}
\ee
In the fermionic sector, we consider the following invariant terms:
\bea
\LL^b_1&=&\left(-\frac{i}{2}\right)
{\bar b}^W_L \gamma_\mu b^W_L tr(T V_\mu)\nn\\
\LL^b_2&=&
\frac{G_F}{\sqrt{2}}({\bar b}^W_L \gamma_\mu b^W_L)^2
\eea
whose contribution to $\Delta\LL$ is given by:
\be
(\Delta{\cal L})_b=\beta_1 \LL^b_1
+ \beta_2 \LL^b_2
\ee
The coefficients $a_0, a_1, a_8, \beta_1, \beta_2$
are easily found by comparing $(\Delta\LL)_B$, $(\Delta\LL)_b$
with the effective hamiltonians $H(\Delta B=2)$ of eq. (3.7),
$H_{NC}$ of eq. (3.29) and $H_b$ of eq. (3.42)
\footnote{The contribution of $\LL_{cl}$ to the right-hand side
of the eq. (4.1) vanishes when we extract the leading top effects.
Indeed $\LL_{cl}$ gives a divergent contribution to the relevant
Green functions, which we choose to subtract at vanishing external
momenta, as an additional prescription to deal with the new infinities
of the effective non-renormalizable theory.}.
To do this one should expand the various combinations appearing
in the expressions of the invariants $\LL_i$. For instance:
\be
tr(T V_\mu)=i \frac {g}{\cos\theta} Z_\mu + \frac{2 i}{v} \partial_\mu
\xi^3+...
\ee
One has:
\bea
a_0\LL_0&=&-\dd\frac{1}{4}a_0 v^2(g W^3_\mu-g'B_\mu)^2+...\nn\\
a_1\LL_1&=& \dd\frac{1}{2} a_1 gg'B_{\mu\nu}(\dmua {W^3}^\nu-\partial^\nu
{W^3}^\mu)+...\nn\\
a_8\LL_8&=&-\dd\frac{1}{4} a_8 g^2(\dmu {W^3}_\nu-\partial_\nu
{W^3}_\mu)^2+...
%\nn\\
%a_{13}\LL_{13}&=& -\dd\frac{1}{2} a_{13}\dmu (g {W^3}_\nu-g'B_\nu)\cdot
%\dmua (g {W^3}^\nu-g'B^\nu)+...\nn\\
\eea
The dots stand for trilinear and quadrilinear terms in
the gauge vector bosons and for terms containing
the would-be Goldstone bosons, needed to ensure the gauge invariance
of each structure.
By evaluating the contribution of the above terms
to the vacuum polarization functions one can relate the
$a_i$ coefficients to the $\epsilon$ parameters as follows
\footnote{$\epsilon_2$ and $\epsilon_3$ receive also an additional
contribution from the operator $\LL_{13}$ (see ref. \cite{par}), which however
can be eliminated by using the equations of motion.}:
\bea
\epsilon_1&=&2 a_0\nn\\
\epsilon_2&=&-g^2 a_8\nn\\
\epsilon_3&=&-g^2 a_1
%\epsilon_2&=&-g^2 (a_8+a_{13})\nn\\
%\epsilon_3&=&-g^2 (a_1+a_{13})
\eea
{}From the behaviour of the $\epsilon$'s in the large $m_t$ limit,
given in eq. (3.32-34), we find:
\bea
a_0&=&\frac{1}{2} (\Delta\rho)_{top} +...                 \nn\\
%(a_8+a_{13})&=&\frac{1}{16\pi^2}\left[ln(\frac{m_t}{M_Z})\right]+...\nn\\
%(a_1+a_{13})&=&\frac{1}{16\pi^2}\left[\frac{1}{3}
a_8&=&\frac{1}{16\pi^2}\left[ln(\frac{m_t}{M_Z})\right]+...\nn\\
a_1&=&\frac{1}{16\pi^2}\left[\frac{1}{3}
             ln(\frac{m_t}{M_Z})\right]+...
\eea
where we have defined:
\be
(\Delta\rho)_{top} = \frac{3 G_F m_t^2} {8 \pi^2 \sqrt{2}}
\ee
Similarly, in the fermionic sector one finds:
\bea
\beta_2&=&-\frac{1}{3} (\Delta\rho)_{top} \left[ 4 f(\frac{m_t^2}{M_W^2})
                                          \eta \right]
\nn\\
\beta_1&=&-\frac{2}{3} (\Delta\rho)_{top}
\eea
The term in square brackets in the left-hand side of eq. (4.12)
is equal to one in the large $m_t$ limit and for QCD interactions turned off.

So far we have just recasted the content of the section 3
into a more elegant form, which however does not seem to provide any
additional information with respect to what already seen in the separate
discussion of the various physical effects.
To appreciate the usefulness of the point of view adopted
here we will mention two facts.

As stressed in the
second section, the violation of the decoupling property
is related to a hierarchy of coupling constants which
may arise by considering the low energy limit of a
given fundamental theory. In the case of an heavy top
quark, such hierarchy, in the ideal case, is represented
by the inequality:
\be
y_l,~ g,~ g'\ll y_t
\ee
where $y_l$ and $y_t$ are the Yukawa couplings for the light quarks
and top quark, respectively. If we consider the extreme case
when all the coupling constants but the top one are put to zero,
we are lead to conclude that for the top quark the violation of the decoupling
property is modeled by a pure Yukawa interaction.
This point is particularly transparent in the effective lagrangian
we have obtained. In $\LL_{eff}$ the coefficients $a_0$,
$\beta_1$ and $\beta_2$ represent the leading effects in the large
$m_t$ limit. If we turn the gauge interactions off, such effects
do not collapse. Indeed the operators $\LL_0$, $\LL^b_1$
and $\LL^b_2$ do not vanish, indicating the Yukawa origin in the SM
of the largest corrections due to the top quark
\footnote{
More generally, in the gaugeless limit we may regard the gauge vector
bosons appearing in the various operators as classical external fields
coupled to light fermions, Higgs and Goldstone bosons, which are
the quantum degrees of freedom in the surviving Yukawa theory.}.

Second, on the practical side, the effective lagrangian
can be seen as the book-keeping of an infinite set of Ward
identities which may be useful in actual computations.
Rather than analyzing the general structure to these
identities we will discuss their physical content on one
example.
Suppose we are interested in the evaluation of the
coefficient $\beta_1$ of $\LL_{eff}$, which is related to the
$Z b {\bar b}$ vertex correction. We should compute the
contribution of the top quark to the operator $\LL^b_1$.
By expanding the exponential of the would-be Goldstone
fields in the combination $tr(T V_\mu)$ (see eq. (4.7)), one obtains:
\be
\LL^b_1=-\frac{i}{2}
{\bar b}^W_L \gamma_\mu b^W_L \left[
i \frac {g}{\cos\theta} Z_\mu + \frac{2 i}{v} \partial_\mu \xi^3+...
\right]
\ee
This equation show that $SU(2)_L \otimes U(1)_Y$ gauge invariance
relates the $Z b {\bar b}$ function to the $\xi^3 b {\bar b}$
function in a well precise way and that, to compute $\beta_1$,
we can in fact consider the latter, by retaining the term linear
in the $\xi^3$ momentum.

It is clear that one does not need the effective lagrangian $\LL_{eff}$ to
derive the Ward Identities of $SU(2)_L \otimes U(1)_Y$, which
are implied just by the symmetry and the particle content.
It is however true that many of these identities can be
in practice read immediately from $\LL_{eff}$, with no
further effort.
In this sense,
the situation closely resembles to what one had with the current
algebra (whose analogue here is $SU(2)_L \otimes U(1)_Y$) and
the PCAC hypothesis (the spontaneous breaking of the symmetry)
in the old times. Indeed they can be either analyzed in abstract
or, as happened with the non-linear $\sigma$-model,
in the context of a specific field theoretical realization.
Each of the two possibilities has its own advantages and can be preferred
depending on the specific problem at hand.

In the last part of this section we will show how to compute the
coefficients $a_0$, $\beta_1$ and $\beta_2$ exploiting the
relevant Ward identities  and the underlying Yukawa nature of the
effects.
We start from the definition of the gaugeless limit of the SM in the top-bottom
sector, assuming a vanishing mass for the bottom:
\bea
\LL_Y&=&
i {\bar b}\gamma^\mu \partial_\mu b+
i {\bar t}\gamma^\mu \partial_\mu t\nn\\
&+&\frac{v^2}{4}tr(\partial_\mu U^\dagger \partial^\mu U)\nn\\
&-&\left(\bar t_L \bar b^W_L\right) U
\left(\begin{array}{cc}m_t&0\\ 0&0
\end{array}\right)
\left(\begin{array}{c}t_R\\b^W_R\end{array}\right) + ~h.c.+...
\eea
The dots stand for additional terms as, for instance, those depending
on the Higgs field. From the lagrangian $\LL_Y$ in eq. (4.15),
the following Feynman rules are derived:
\bea
\xi^+ {\bar t} b &\leftrightarrow& -\frac{\sqrt{2}}{v} m_t a_-\nn\\
\xi^- {\bar b} t &\leftrightarrow& \frac{\sqrt{2}}{v} m_t a_+\nn\\
\xi^3 {\bar t} t &\leftrightarrow& \frac{m_t}{v} (a_+ - a_-)\nn\\
\xi^+ {\bar b} b &\leftrightarrow& 0
\eea
where
\be
a_\pm=\frac{1\pm \gamma_5}{2}
\ee

\noindent
$\bullet$
{\bf $B^0-{\bar B^0}$ oscillations}

To compute $\beta_2$ we consider the top quark contribution to the four-fermion
operator $\LL^b_2$. In the Yukawa theory defined above by $\LL_Y$ such
contribution is represented by two independent box diagrams,
with top and charged Goldstone bosons circulating in the loop, which we will
evaluate taking vanishing external momenta. The total amplitude $A_{box}$,
ultraviolet convergent, is given by:
\bea
A_{box}&=&
\int \frac{d^4 k}{(2\pi)^4} {\bar u}(0) (\frac{\sqrt{2}}{v} m_t a_+)
     \frac{i}{\slash k - m_t} (-\frac{\sqrt{2}}{v} m_t a_-) u(0)
     \frac{i}{k^2}\nn\\& & \cdot
     {\bar v}(0) (\frac{\sqrt{2}}{v} m_t a_+)
     \frac{i}{\slash k - m_t} (-\frac{\sqrt{2}}{v} m_t a_-) v(0)
     \frac{i}{k^2}+~~crossed\nn\\
&=&-\frac{i}{v^2}\frac{(\Delta\rho)_{top}}{3}
\bigl[{\bar u}(0) \gamma^\mu a_- u(0)\cdot {\bar v}(0) \gamma^\mu a_- v(0)\nn\\
& &+{\bar v}(0) \gamma^\mu a_- u(0)\cdot {\bar u}(0) \gamma^\mu a_- v(0)\bigr]
\eea
{}From comparison with eq. (4.5-6), one has:
\be
\beta_2=-\frac{1}{3}(\Delta\rho)_{top}
\ee
which is the correct large $m_t$ limit of $\beta_2$ in eq. (4.12).

\noindent
$\bullet$
{\bf $Z{\bar b} b$ vertex corrections}\cite{Zbb}

{}From eqs. (4.6) and (4.7) one has:
\be
\beta_1 \LL^b_1=\beta_1
\left[
\frac {g}{2 \cos\theta} Z_\mu + \frac{1}{2 v} \partial_\mu \xi^3+...
\right]
{\bar b}^W_L \gamma_\mu b^W_L
\ee
We choose to compute the correction to the $\xi^3 {\bar b} b$ Green function.
{}From the second term in the previous equation, we derive the amplitude:
\be
A_{\xi^3 {\bar b} b}=
\beta_1 {\bar u}(p)
\frac{\slash p}{v} a_- v(0)
\ee
where $p_\mu$ is the four-momentum of the incoming $\xi^3$.
We should compare this result with the (finite) contribution due to
the top-$\xi$ loop, which reads:
\bea
A_{\xi^3 {\bar b} b}&=&
\int \frac{d^4 k}{(2\pi)^4} {\bar u}(p) (\frac{\sqrt{2}}{v} m_t a_+)
     \frac{i}{\slash k + \slash p - m_t} \frac{m_t}{v} (a_+- a_-)\nn\\
&&\cdot   \frac{i}{\slash k - m_t} (-\frac{\sqrt{2}}{v} m_t a_-) v(0)
     \frac{i}{k^2}\nn\\
&=&
-2i \frac{m_t^2}{v^2} \int \frac{d^4 k}{(2\pi)^4}
\frac{m_t^2}{k^2 (k^2 -m_t^2)^2} \cdot {\bar u}(p)
\frac{\slash p}{v} a_- v(0)+...\nn\\
&=&-\frac{2}{3}(\Delta\rho)_{top}
\cdot {\bar u}(p) \frac{\slash p}{v} a_- v(0)
\eea
where dots in the second equality stand for higher-order terms in $p$.
{}From eqs. (4.21) and (4.22) one finds:
\be
\beta_1=-\frac{2}{3} (\Delta\rho)_{top}
\ee
in agreement with the result given in eq. (4.12).

\noindent
$\bullet$
{\bf $\Delta\rho$ parameter}\cite{rho}

To evaluate the $a_0$ coefficient, we look at the expansion of the
operator $a_0 \LL_0$:
\bea
a_0 \LL_0&=&a_0\dd\frac{v^2}{4}[tr(TV_\mu)]^2\nn\\
&=&a_0\left[\frac{1}{4}\frac{g^2 v^2}{\cos^2\theta} Z^\mu Z_\mu+
...+\partial^\mu \xi^3 \partial_\mu \xi^3+...\right]
\eea
{}From this expression
we notice that $a_0$ represents also
a wave function renormalization of the $\xi^3$ field.
Thus we are lead to consider the two-point function
$-i \Pi_{\xi^3\xi^3}(p^2)$
for the field $\xi^3$. To match the results in the fundamental theory
and in the effective one, we have to impose:
\be
\Pi_{\xi^3\xi^3}^{tt}(p^2)+\Pi_{\xi^3\xi^3}^C(p^2)=\Pi_{\xi^3\xi^3}^0(p^2)
\ee
where $\Pi_{\xi^3\xi^3}^{tt,C,0}(p^2)$ are, respectively, the contribution
of the top quark loop, the contribution of the counterterm and the
contribution of $a_0 \LL_0$.
The counterterm needed to cancel the divergences of
$\Pi_{\xi^3\xi^3}^{tt}(p^2)$
is given by:
\be
\frac{\delta v^2}{4}tr(\partial_\mu U^\dagger \partial^\mu U)
=\frac{1}{2}\frac{\delta v^2}{v^2}\partial^\mu \xi^i \partial_\mu \xi^i+...
\ee
{}From this term we derive:
\be
\Pi_{\xi^3\xi^3}^{C}(p^2)=\frac{\delta v^2}{v^2} p^2
\ee
On the other hand from eq. (4.24), one has:
\be
\Pi_{\xi^3\xi^3}^{0}(p^2)=2 a_0 p^2
\ee
Before computing the top quark loop, we observe that we can get rid of the
counterterm contribution by writing the analogous matching condition for
the $\xi^+\xi^+$ two-point function, $\Pi_{\xi^+\xi^+}(p^2)$:
\be
\Pi_{\xi^+\xi^+}^{tb}(p^2)+\Pi_{\xi^+\xi^+}^C(p^2)=\Pi_{\xi^+\xi^+}^0(p^2)
\ee
{}From eqs. (4.24) and (4.26), we obtain:
\be
\Pi_{\xi^+\xi^+}^{C}(p^2)=\frac{\delta v^2}{v^2} p^2
\ee
and
\be
\Pi_{\xi^+\xi^+}^{0}(p^2)=0
\ee
By combining the conditions (4.25) and (4.29) and by making use of
eqs. (4.27-28) and (4.30-31), we find:
\be
2 a_0 p^2=\Pi_{\xi^3\xi^3}^{tt}(p^2)-\Pi_{\xi^+\xi^+}^{tb}(p^2)
\ee
where it is understood that in the right-hand side we have to
consider only the contribution proportional to $p^2$.
By a direct evaluation of the Feynman amplitudes one obtains:
\be
-i \Pi_{\xi^3\xi^3}^{tt}(p^2)=
12 (\frac{m_t}{v})^2 p^2
\int \frac{d^d k}{(2\pi)^4} \frac
{\left[(1-\frac{2}{d}) k^4 -2 m_t^2(1-\frac{1}{d}) k^2+m_t^4\right]}
{(k^2-m_t^2)^4}+...
\ee
and
\be
- i \Pi_{\xi^+\xi^+}^{tb}(p^2)=
12 (\frac{m_t}{v})^2 p^2
\int \frac{d^d k}{(2\pi)^4} \frac
{(1-\frac{2}{d}) }
{k^2 (k^2-m_t^2)}+...
\ee
where dots stand for higher terms in the $p^2$ expansion.
The equation (4.32) now reads:
\be
-2i a_0 p^2=
-6 (\frac{m_t}{v})^2 p^2
\int \frac{d^4 k}{(2\pi)^4} \frac
{m_t^4 }
{k^2 (k^2-m_t^2)^3}
\ee
from which we obtain:
\be
2 a_0(=\epsilon_1)=(\Delta\rho)_{top}
\ee
in agreement with eqs. (4.10).

These examples show how the violation of the decoupling theory in
the SM with an heavy top quark is related to the underlying Yukawa
theory. From the practical point of view, one may have the impression
of an unnecessary complication in dealing with a simple 1-loop computation.
Moreover, to reach the accuracy required to compare the experimental
prediction to the theoretical expectation, one should also include the
corrections to the Yukawa limit taken in the fundamental theory.

Nevertheless
%, beyond the advantage of clarifying some conceptual issue,
the strategy followed above has been already useful
in attacking more challenging computations as, for example,
those concerning the leading two-loop effects in the pure
electroweak theory \cite{m4} (O($G_F^2 m_t^4$)),
the mixed strong and electroweak corrections of O($\alpha_s G_F m_t^2$)
\cite{asm2} and the non-leading corrections - O($G_F^2 M_W^2 m_t^2$) -
to the $\rho$ parameter \cite{nonlead}.

We conclude this section with a comment concerning the size
of the corrections to the results obtained in the gaugeless limit
of the SM. These corrections are of order $(M_W/m_t)^2$ and, in principle,
they can be large compared to the leading order results.
At one-loop order one has two extreme cases. One-loop two-point functions
are essentially untouched by the gaugeless limit, the only
approximation coming from the subsequent expansion in $p^2/m_t^2$,
which however works remarkably well already for $m_t^2\simeq 2 p^2$.
On the other hand, vertex and box corrections may
be largely modified in the full gauge theory. Consider for instance the
function $f(m_t^2/M_W^2)$ appearing in the evaluation of the box diagram
for $B^0-{\bar B^0}$ oscillation (see eq. (3.7)).
The asymptotic value $f(\infty)=1/4$ is not so close to the more realistic
case $f(4)\simeq 0.57$. Moreover, also the first term
in the expansion of $f(1/y)$ around $y=0$ fails to provide the
right correction (it does not even give the correct sign!):
\be
f(\frac{1}{y})=\frac{1}{4}-(\frac{9}{4}+\frac{3}{2} ln y) y+...
\ee
By truncating the expression above at first order in $y$, one obtains
$f(4)=0.25-0.043=0.207$ rather far away from the physical value.

\resection{Heavy Fermions and Chiral Anomalies}

In the previous section we have imposed the $SU(2)_L\otimes U(1)_Y$ gauge
invariance on the low-energy lagrangian
$\LL_{eff}$.
The physical basis of this requirement is the fact that the heaviness
of the top quark is achieved with a gauge invariant procedure,
adjusting the magnitude of its Yukawa coupling to be
(much) bigger than the other coupling constants in the theory.
This accidental hierarchy will always respect the gauge invariance
of the theory which has to persist also in the low-energy approximation.
There is however a subtlety in the mechanism which maintains
gauge invariance, due to the features of the
classical term $\LL_{cl}$ in $\LL_{eff}$. The light matter fields
entering $\LL_{cl}$ do not form an anomaly-free set of
chiral fermions. This means that at one-loop order the gauge currents
are not conserved and the $O(\slash h)$ contribution of $\LL_{cl}$
to the effective action is not gauge invariant. On the other hand,
since the total effective action must be gauge invariant, the gauge variation
of the terms induced by $\LL_{cl}$ has to be exactly compensated by
the gauge variation of $\Delta\LL$.
So far we have only included gauge invariant operators in $\Delta\LL$.
In this section we will identify the additional, non-invariant
contributions in $\Delta\LL$ and we will detail the mechanism of
anomaly cancellation.

We recall that anomalies may occur as violations of symmetry properties
of a classical theory, in the regularization procedure which
underlies the construction  of the corresponding quantum theory \cite{Anomaly}.
The classical example is given by the axial current in QED (see eq. (2.1)):
\be
j^\mu_5={\bar \psi} \gamma^\mu\gamma_5 \psi
\ee
At the classical level, the divergence of $j^\mu_5$ is proportional to the
pseudoscalar density:
\be
\partial_\mu j^\mu_5=2 i M {\bar \psi} \gamma_5 \psi
\ee
so that, for $M=0$, $j^\mu_5$ is conserved. Indeed, for $M=0$ the QED
lagrangian
of eq. (2.1) possesses the chiral symmetry $U(1)_L\otimes U(1)_R$,
which leads to the separate conservation of the vector current and of the
axial-vector one.
It is well known that this is no more true at one-loop order and
the Ward identity (5.2) is replaced by:
\be
\partial_\mu j^\mu_5=2 i M {\bar \psi} \gamma_5 \psi
+\frac{e^2}{16\pi^2} F_{\mu\nu} {\tilde F}^{\mu\nu}
\ee
where ${\tilde F}^{\mu\nu}=1/2\epsilon^{\mu\nu\rho\sigma} F_{\rho\sigma}$.
On the other hand, the vector current, associated to the $U(1)$ local
invariance, is still conserved.

In a chiral gauge theory, with left and right-handed
fermions transforming
according to inequivalent representations of the gauge group, there
will be both vector and axial-vector gauge currents. In this case
the problem of a possible (and unacceptable) breaking of the gauge
symmetry via
quantum effects immediately arises. It turns out that in this case
it is the fermion content of the theory that decides if the gauge
currents are anomalous or not, and a simple criterion can be
formulated.
It is convenient to define all the fermion fields to be left-handed.
This is always possible: whenever a right-handed field $\psi_R$ occurs,
it may be always replaced by its charge-conjugate left-handed counterpart
$(\psi^c)_L$, with $\psi^c = C{\bar\psi}^T$.
In this way all the gauge currents are of the kind:
\be
j^A_\mu={\overline{\Psi_L}}\gamma_\mu T^A \Psi_L
\ee
where $\Psi_L$ stands for the collection of fermion fields and $T^A$ is
the set of gauge generators for the representation $\Psi_L$.
An anomaly-free theory is characterized by the condition:
\be
D^{ABC}=tr(T^A\{T^B,T^C\})=0
\ee
corresponding to the vanishing of all possible quantum contributions
to the anomalies via triangle diagrams.
For instance, in the SM, the above condition is equivalent to
the requirement:
\be
tr(Q_{em})=0
\ee
the trace being performed over all $SU(2)$ doublets. A full
fermion generation satisfies eq. (5.6), since the quark contribution,
$3\times (2/3-1/3)=+1$ exactly compensates the leptonic one, $-1$.
An immediate consequence is that the removal of the top quark from
the low-energy spectrum of $\LL_{cl}$ makes the theory anomalous.

Instead of investigating the problem in the SM, we consider here
a simplified model, based on an abelian gauge symmetry $U(1)$.
The matter content of the theory consists of a "lepton" $l$ and a
"quark" $q$ whose left-handed component transforms according
opposite $U(1)$ charges. The right-handed components are taken
invariant:
\bea
l'_L&=&\dd e^{i\dd\alpha(x)} l_L\nn\\
q'_L&=&\dd e^{-i\dd\alpha(x)} q_L\nn\\
l'_R&=&0\nn\\
q'_R&=&0
\eea
To trigger the spontaneous breaking of the gauge symmetry we
introduce also a complex scalar field $\varphi$ transforming
as follows:
\be
\varphi'=\dd e^{i\dd\alpha(x)} \varphi
\ee
Finally we provide the equivalent of a lepton number $L$ and
a baryon number $B$, by requiring invariance under a global
$U(1)_L\otimes U(1)_B$ symmetry, with natural assignment:
\bea
L(q)&=L(\varphi)=0~~~~~~~~L(l)&=1\nn\\
B(l)&=B(\varphi)=0~~~~~~~~B(q)&=1
\eea
The lagrangian for this model reads:
\bea
\LL&=&-\frac{1}{4} F_{\mu\nu} F^{\mu\nu}\nn\\
&+&i {\bar {l_L}} \gamma^\mu (\partial_\mu - ig A_\mu) l_L
+i {\bar {l_R}} \gamma^\mu \partial_\mu l_R\nn\\
&+&i {\bar {q_L}} \gamma^\mu (\partial_\mu + ig A_\mu) q_L
+i {\bar {q_R}} \gamma^\mu \partial_\mu q_R\nn\\
&+&D_\mu\varphi^\dagger D^\mu\varphi-V(\varphi^\dagger \varphi)\nn\\
&-&y_l({\bar {l_L}}\varphi l_R + h.c.)
-y_q({\bar {q_L}}\varphi q_R + h.c.)
\eea
The potential $V(\varphi^\dagger \varphi)$ gives rise
to the spontaneously broken phase, if we make the usual choice:
\be
V(\varphi^\dagger \varphi)=\mu^2 \varphi^\dagger \varphi
+\lambda (\varphi^\dagger \varphi)^2
\ee
with $\mu^2<0$ and $\lambda>0$.
The minimum is at:
\be
<\varphi>=\frac{v}{\sqrt{2}}
\ee
with
\be
v^2=-\frac{\mu^2}{\lambda}
\ee
We shift the scalar field as follows:
\be
\varphi=\frac{(\sigma+v)}{\sqrt{2}}\dd e^{i\dd\frac{\xi}{v}}
\ee
Notice that the would-be Goldstone boson $\xi$ undergoes the
following gauge transformation:
\be
\xi'=\xi+\alpha(x) v
\ee
The gauge symmetry is spontaneously broken and all the particles
become massive via the Higgs mechanism. The mass spectrum is the following:
\bea
M_A^2&=g^2 v^2\nn\\
M_\sigma^2&=2 \lambda v^2\nn\\
m_l&=\dd\frac{y_l v}{\sqrt{2}}\nn\\
m_q&=\dd\frac{y_q v}{\sqrt{2}}
\eea
We are interested in the gauge current $j^\mu$, given by:
\bea
j^\mu&=&{\bar {l_L}}\gamma^\mu l_L-{\bar {q_L}}\gamma^\mu q_L
      + i(\varphi^\dagger D^\mu \varphi-(D^\mu \varphi^\dagger) \varphi)\nn\\
&=&{\bar {l_L}}\gamma^\mu l_L-{\bar {q_L}}\gamma^\mu q_L-v\partial^\mu \xi+...
\eea
In the second equality we have used the parametrization for
the scalar field given in eq. (5.14), writing down explicitly only the term
linear in the field $\xi$. Dots denote terms with $\sigma$ or
more than one $\xi$, which, for the sake of simplicity, we will neglect
from now on.
The divergence of $j_\mu$ reads:
\be
\partial^\mu j_\mu=-i m_l {\bar l}\gamma_5 l +i m_q {\bar q} \gamma_5 q
-v \Box\xi+...
\ee
By taking into account the equations of motion for the field $\xi$:
\be
\Box \xi=
-\frac{i}{v} m_l {\bar l}\gamma_5 l +\frac{i}{v} m_q {\bar q} \gamma_5 q+...
\ee
we find that the gauge current is conserved also in the spontaneously
broken phase, at least at the classical level.
To see what happens with the quantum corrections, we write the generator $T$
of the gauge transformations (5.7) in the base $l_L,q_L,(l^c)_L,(q^c)_L$:
\be
T=\left(
\begin{array}{cccc}
1 & 0  & 0 & 0\\
0 & -1 & 0 & 0\\
0 & 0  & 0 & 0\\
0 & 0  & 0 & 0
\end{array}
\right)
\ee
The condition (5.5) for the absence of anomalies in the gauge currents
now reads $tr(T^3)=0$, which is clearly satisfied by $T$.
In terms of triangle diagrams, the quark contribution is exactly
cancelled by the lepton contribution, as in the SM.
The model is anomaly-free.

To mimic the case of the SM, we now assume that the quark is much heavier
than the other particles. As one can see from eq. (5.16), this means
that we are postulating the following hierarchy:
\be
y_q\gg y_l,g,\lambda
\ee
Such a choice does not interfere with the gauge invariance of the model,
which we have checked above. As in the previous section, we introduce
a low-energy effective action given by:
\be
S_{eff}=S_{cl}+\Delta S
\ee
Here $S_{cl}$ is obtained from the original lagrangian $\LL$ of
eq. (5.10) simply
by dropping the terms containing the heavy field $q$.
The theory described by $S_{cl}$ is still formally gauge invariant.
The gauge current $j^\mu_{cl}$ is given by:
\bea
j^\mu_{cl}&=&{\bar {l_L}}\gamma^\mu l_L
      + i(\varphi^\dagger D^\mu \varphi-(D^\mu \varphi^\dagger) \varphi)\nn\\
&=&{\bar {l_L}}\gamma^\mu l_L-v\partial^\mu \xi+...
\eea
with vanishing classical divergence. However, due to the anomalous
fermion content of the theory, quantum corrections modify the classical
Ward identity, and, as in the case of the axial vector current in QED,
one obtains
\be
\partial_\mu j^\mu_{cl}=\frac{1}{6}
\frac{g^2}{16\pi^2} F_{\mu\nu} {\tilde F}^{\mu\nu}=G(x)
\ee
%where we have neglected the lepton mass $m_l$.
This result immediately
implies the breaking of gauge invariance.
We consider the gauge variation of $S_{cl}$. We obtain:
\bea
\delta \left[S_{cl}\right] &=& - \int dx \alpha(x) \partial_\mu j^\mu_{cl}\nn\\
&=& - \int dx \alpha(x) G(x)\nn\\
&\neq& 0
\eea
where we have denoted by $[S_{cl}]$ the full one-loop effective action
induced by $S_{cl}$.

There is not much freedom to repair this situation. The only possibility
is that $\Delta S$ in eq. (5.22), which is a genuine $O(\slash h)$ term,
has a gauge variation which exactly compensates the one given in eq. (5.25),
namely:
\be
\delta (\Delta S)=\int dx \alpha(x) G(x)
\ee
Indeed, we may try to find the general - $CP$ invariant - solution
to the equation (5.26). To this end it is useful to split $\Delta S$
into a parity violating part $\Delta S_{PV}$ and a parity
conserving term $\Delta S_{PC}$. Indeed it is not restrictive
to require that $\Delta S_{PC}$ is gauge invariant so that it does
not contribute to the previous equation. The term $\Delta S_{PC}$,
analogous to the term $(\Delta\LL)_B$ of eq. (4.4), can be
determined via suitable matching conditions, as explained in section 4,
but is irrelevant to the present discussion.
The general solution to eq. (5.26) has now the form:
\be
\Delta S_{PV}=\Delta S_{PV}^0+\Delta S_{PV}^1
\ee
where $\Delta S_{PV}^0$ is the general solution of the homogeneous
equation $\delta(\Delta S_{PV}^0)=0$, i.e. the set of all possible
gauge and $CP$ invariant, $P$ violating operators.
Such operators do not exist.
\footnote{We are using the possible assignment:
$P=+1$ and $C=-1$, for the field $\xi$.}.
So we remain with $\Delta S_{PV}^1$, a particular solution
of eq. (5.26).
Long ago Wess and Zumino found
the solution \cite{WZ}:
\be
\Delta S_{PV}^1=\frac{1}{v} \int dx~ \xi(x) G(x)
\ee
which indeed satisfies eq. (5.26), as can be seen by using the transformation
properties of $\xi$ in eq. (5.15) and the invariance of $G$.

The final gauge invariant effective action is given by:
\be
S_{eff}=S_{cl}+\Delta S_{PC} + \frac{1}{6 v}
\int dx~ \xi F_{\mu\nu} {\tilde F^{\mu\nu}}
\ee
Notice that $S_{eff}$ is non-renormalizable, as in fact it should be,
since otherwise, we would have "integrated away" the anomaly.
The non-renormalizability is related to the restricted
domain of applicability
of the effective theory. Such domain is bounded in energy by
some critical value $E_c$, beyond which the breakdown of
perturbative unitarity signals the inadequacy of $S_{eff}$ to approximate
the full theory.

A similar mechanism of anomaly cancellation is active
in the low energy effective lagrangian
$\LL_{eff}$ of the previous section, obtained from the SM
in the heavy $m_t$ limit \cite{FMM}. In this case we have to deal with
the additional constraint given by the lightness of the bottom
quark, which belongs to the low-energy part of the spectrum.
At first sight this new element seems to lead to a
contradictory situation. On one hand the consistency
conditions which the $SU(2)_L\otimes U(1)_Y$ anomalies
must satisfy require for the bottom contribution to the anomaly
to be fully included in the Wess-Zumino term $\Delta S$.
On the other hand, being the bottom a light field, nothing
prevents the separate evaluation of the bottom contribution
to the modified Ward identities in $S_{cl}$. It is a property
of the non-linear realization of the $SU(2)_L\otimes U(1)_Y$
symmetry  that makes the latter vanish, solving the
paradox.

Similar mechanisms take also place in supersymmetric extensions
of the SM analyzed in the large $m_t$ limit \cite{Porrati}.

To conclude this section we illustrate the physical relevance
of the Wess-Zumino term by discussing the cancellation
of the gauge dependence in the scattering amplitude for
$l {\bar l} -> A A$.
This amplitude, beyond the tree-level contributions,
receives from the lepton loop three independent one-loop corrections:
the exchange $A_V$ of the vector boson $A$ in the $s$-channel,
the exchange $A_\xi$ of a would-be
Goldstone boson $\xi$ and finally
the contribution $A_{WZ}$ of the Wess-Zumino term through the exchange
of $\xi$.
In a generic $R_\lambda$ gauge ($\lambda$ denoting the gauge parameter)
we have:
\be
A_V={\bar v} (i g \gamma^\mu a_-) u \cdot
    \frac{-i}{p^2-M_A^2} \cdot
    \left[ g_{\mu\nu} -\frac{1-\lambda^{-1}}{p^2-\frac{M_A^2}{\lambda}}
    p_\mu p_\nu \right] <ig j^\nu>
\ee
where $<ig j_\mu>$ denotes the insertion of the
leptonic current between two vector bosons via the lepton loop.
{}From the anomalous Ward identity one obtains:
\be
i p^\mu <j_\mu> = -i m_l <{\bar l} \gamma_5 l> + <G>
\ee
so that one finds:
\be
A_V=A_V^1+A_V^2
\ee
with
\be
A_V^1={\bar v} (i g \gamma^\mu a_-) u \cdot
    \frac{-i}{p^2-M_A^2} \cdot
    <ig j_\mu>
\ee
and
\be
A_V^2= - g^2 m_l ({\bar v}\gamma_5 u) \cdot
     \frac{1}{p^2-M_A^2} \cdot
     \frac{1-\lambda^{-1}}{p^2-\frac{M_A^2}{\lambda}}
     \left[ <G>-im_l<{\bar l} \gamma_5 l> \right]
\ee
Notice that $A_V^1$ does not depend on $\lambda$.
The other contributions are given by:
\be
A_\xi= \frac{m_l}{v} ({\bar v}\gamma_5 u) \cdot
     \frac{i}{p^2-\frac{M_A^2}{\lambda}} \cdot
\frac{m_l}{v}<{\bar l} \gamma_5 l>
\ee
and
\be
A_{WZ}= \frac{m_l}{v} ({\bar v}\gamma_5 u) \cdot
     \frac{i}{p^2-\frac{M_A^2}{\lambda}} \cdot
\frac{i}{v}<G>
\ee
Finally the sum of the contributions $A_V^2$, $A_\xi$ and
$A_{WZ}$ is given by:
\be
A_V^2+ A_\xi+A_{WZ}=
-\frac{1}{v^2} m_l ({\bar v}\gamma_5 u) \cdot
\frac{1}{p^2-M_A^2} \left[<G> -i m_l <{\bar l} \gamma_5 l> \right]
\ee
displaying the desired independence on $\lambda$. It is only the
sum of the three contributions that does not depend on $\lambda$.
In particular, neglecting the contribution from the Wess-Zumino
term we would obtain an unacceptable gauge-dependent amplitude.

\vskip 1.truecm

\noindent
{\bf ACKNOWLEDGMENTS}:

I am indebted to L. Maiani and A. Masiero for the very pleasant collaboration
on which most of these lectures are based.
I would like to thank O. Boyarkin, G. Degrassi, S. Rigolin, R. Strocchi
and D. Zeppenfeld for stimulating discussions on the subject of these lectures.
I am grateful to A. Vicini for a critical reading of the manuscript and
many useful suggestions.
A special thank goes to the organizers of the School
R. Manka, J. Polak and M. Zralek as well as to all the participants
for the very nice hospitality enjoyed in Szczyrk, for having provided
an exceptionally good weather and for the enjoyable walks in the Beskidy
mountains.
Finally I would like to thank E. Masso and the Institute de Fisica
de Altas Energias of Autonoma University, D. Espriu
and the Department of Physics of Central University
for their kind hospitality in Barcellona where
these lectures were further reviewed and completed.

\end{document}